# FAST: A Fully-Concurrent Access SRAM Topology for High Row-wise Parallelism Applications Based on Dynamic Shift Operations

Yiming Chen, *Student Member, IEEE*, Yushen Fu, *Student Member, IEEE*, Mingyen Lee, *Student Member, IEEE,* Sumitha George, Yongpan Liu, *Senior Member, IEEE*, Vijaykrishnan Narayanan, *Fellow, IEEE*, Huazhong Yang, *Fellow, IEEE*, and Xueqing Li, *Senior Member, IEEE*

***Abstract*—This paper proposes a fully-concurrent access SRAM topology to handle high-concurrency operations on multiple rows in an SRAM array. Such high-concurrency operations are widely seen in both conventional and emerging applications where high parallelism is preferred, e.g., the table update in a database and the parallel feature update in graph computing. The proposed shift-based parallel access and compute architecture is enabled by integrating the shifter function into each SRAM cell, and by creating a datapath that exploits the high-parallelism of shift operations in multiple rows. An example of a 128-row 16-column shiftable SRAM in 65nm CMOS is designed. Post-layout SPICE simulations show improvements of 5.5x energy efficiency and 27.2x speed in average over a conventional digital near-memory computing scheme. In addition, the design has been fabricated and the measurement results show support of up to 800MHz clock at 1.0V and 1.2GHz at 1.2V.**

## I. Introduction

HIGH-CONCURRENCY access to a structured memory array could be widely seen in many data-intensive applications. Some examples include the table management in database and the weight matrix read and update in graph applications. Conventionally, the read and write access to multiple rows of an embedded memory array is carried out row by row, sequentially. This is due to the sharing of the bitlines and peripheral circuitry for many rows, so as to balance the density and latency well for cache design in most conventional computing tasks. However, in data-intensive high-concurrency memory access applications, it has caused significant latency and become a performance bottleneck, as illustrated in Fig. 1(a). Furthermore, the energy consumption per access is also high due to charging of long bitlines with large parasitic capacitance in read and write operations. Therefore, it is time to re-think the memory access pattern and supporting circuits for emerging high-concurrency applications.

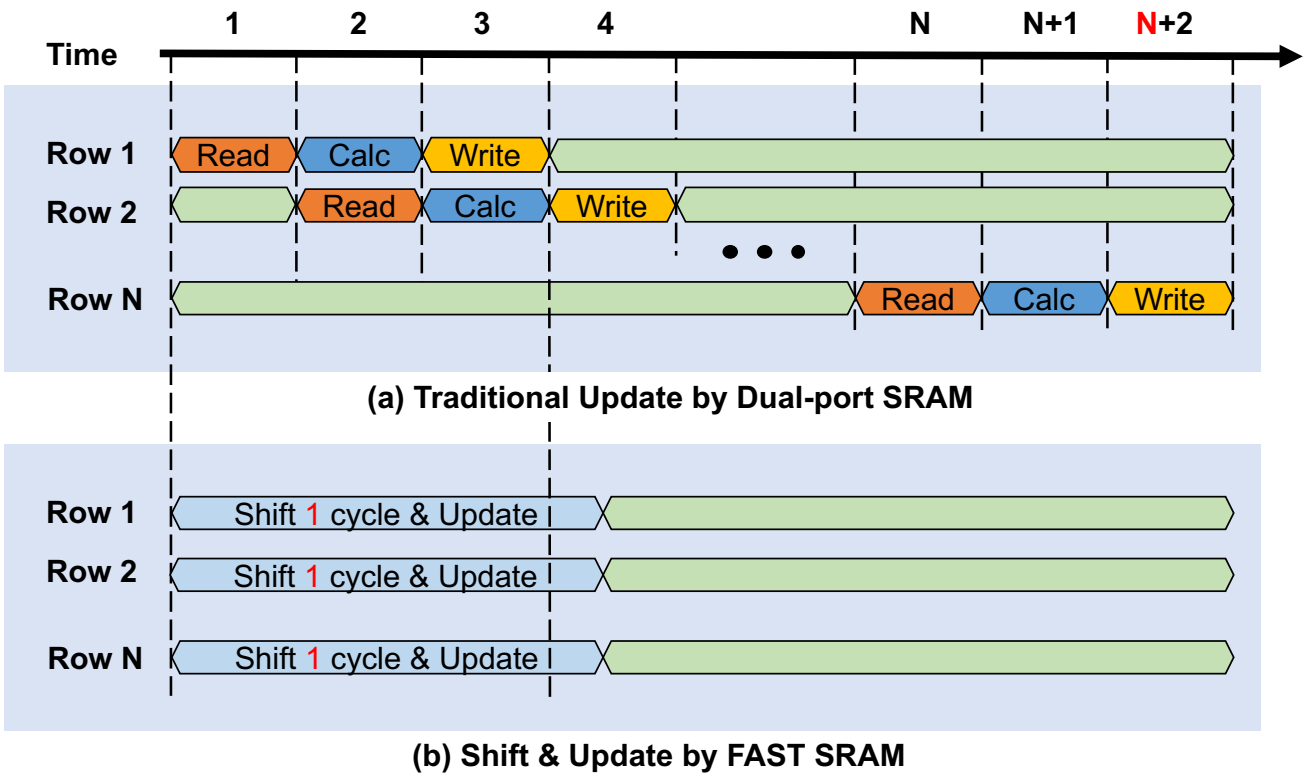

Fig. 1. Comparison between conventional row-by-row data access to a dual-port SRAM in (a), and proposed shift-based *FAST* SRAM supporting full-concurrency parallel read & update in (b).

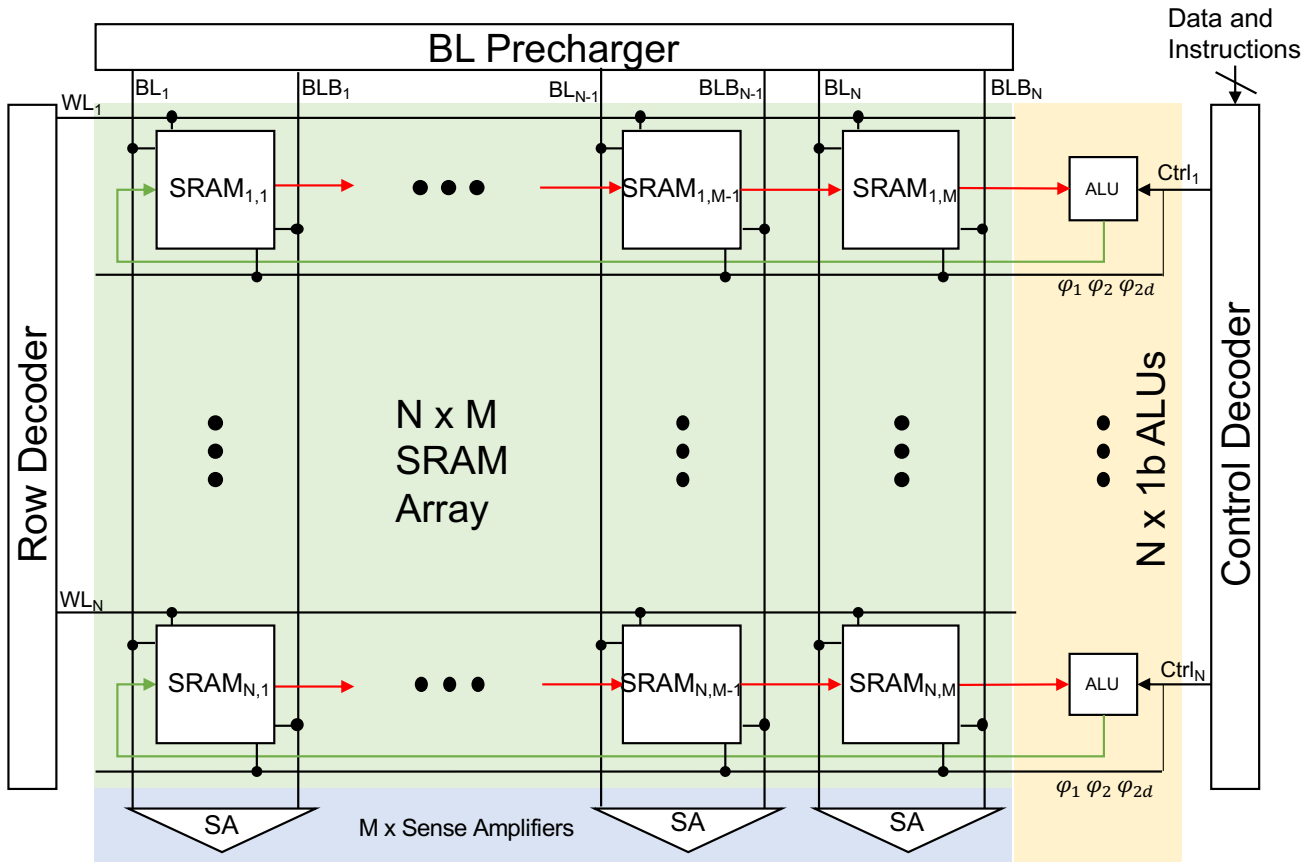

Fig. 2. Shift-based in-memory computing architecture.

There has been an emerging concept of compute-in-memory (CiM) to reduce the data transfer cost between memory and processing units [1][2][3], actually supporting concurrent access to multiple rows. However, existing CiM techniques are limited in the computing functionalities and parallelism. On the one hand, the parallelism of near memory computing style

Date of submission September 5, 2022.

This work was supported in part by NSFC (#61874066, #61934005), and in part by The National Key R&D Program of China (#2018YFA0701500).

Yiming Chen, Yushen Fu, Mingyen Lee, Yongpan Liu, Huazhong Yang, and Xueqing Li are with BNRist/ICFC, The Department of Electronic Engineering, Tsinghua University, Beijing 10084, China (Email: {cym21, fys15, lmy21}@mails.tsinghua.edu.cn, {ypliu, yanghz}@tsinghua.edu.cn, xueqingli@tsinghua.edu.cn).

Sumitha George is with the Department of ECE, North Dakota State University, Fargo, ND, USA (Email: sumitha.george@ndsu.edu).

Vijaykrishnan Narayanan is with the Department of Computer Science and Engineering, Penn State University, University Park, PA 16802 USA (Email: vijay@cse.psu.edu).

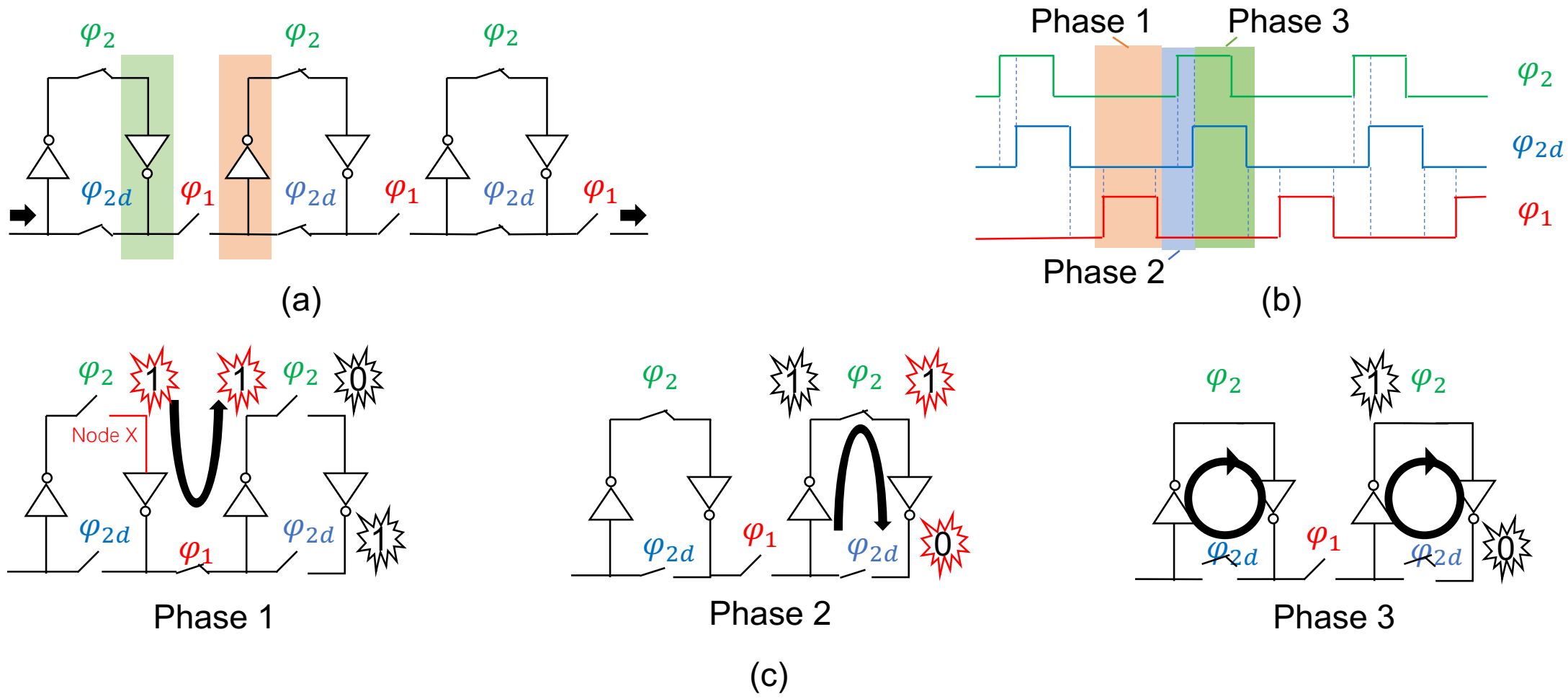

Fig. 3. FAST SRAM: (a) three SRAM in-row shifters in a row (SRAM access transistors not shown); (b) control diagram; (c) shift operation steps.

[4][5][6] is limited by the SRAM data port, which needs high-frequency memory access row by row. On the other hand, high-concurrency memory update, e.g., the feature update in graph computing applications [7][8]), or the delta update of a cache table (and possibly weight update in neural networks [9]), could still be limited by the row-by-row access bottleneck.

To tackle these high-concurrency read and write problems, this paper proposes *FAST*, namely a fully-concurrent access SRAM topology, as shown in Fig. 2. To demonstrate FAST, we take the static random-access memory (SRAM) in the CMOS technology for evaluations of functionality, latency, and energy efficiency. Contributions of this work include:

(i) A dynamic shift-based SRAM capable of reading and writing multiple rows with full row concurrency. The SRAM cell, array, and supporting circuitry are presented;

(ii) A fully-concurrent *in situ* read-compute-and-update architecture supporting a category of parallel computing operations achieved by adding a 1-bit ALU to each row;

(iii) Analysis and evaluations of the proposed techniques based on a showcase of a 128-row 16-column FAST chip implemented in 65nm CMOS, showing significant improvement of speed and energy efficiency over the conventional SRAM solutions.

In the rest of this paper, Section II presents the proposed FAST architecture. Section III evaluates the performance and costs. Section IV concludes this work.

## II. Proposed FAST Memory Architecture and Circuits

This section presents FAST, including the architecture, supporting circuits, and *in situ* computing capabilities. Simulation and evaluation results are provided in Section III.

### *A. Overall Architecture*

The proposed system architecture is shown in Fig. 2. The bitline (BL) precharger and the row decoder are the same as those of a conventional SRAM array. The control decoder serves as an interface to the external processing units such as CPU or FPGA. The SRAM cell in the proposed architecture is designed to support in-cell shift function, so that each row could be cyclically shifted independently (to the right, for example). More circuit details will be introduced subsequently in Section II.B. Based on this, we add a 1-bit arithmetic logic unit (ALU) in each row, connecting the last cell and the first cell. This 1-bit ALU performs 1-bit logic computing, such as 1-bit add. By combining the shift operation and the 1-bit logic operation, multibit operation could be completed naturally in parallel between different rows.

According to the different functions of 1-bit ALU, this scheme could support applications that needs parallel updates to the stored data in several rows. One simple scenario is a high-concurrency access-intensive general cache, such as those of database table and weight matrix during training.

It is noted that there are some emerging SRAM-based CiM schemes to reduce the data transfer costs by enabling in-SRAM computing [10]–[12]. In contrast, this proposed FAST SRAM is different in enabling row-wise *in situ* calculation with direct write-back support. In the following sections, we will show how to implement the multibit addition with cyclic right shift and 1-bit ALU.

### *B. SRAM In-Memory Parallel Shifter*

As mentioned above, the shiftable SRAM design is a key enabler of the parallel search and computing architecture. In order to support in-row shift operation with less area and latency overhead, we propose a shiftable SRAM cell in Fig. 3, including the cell circuit structure in Fig. 3(a), the control flow chart in Fig. 3(b), and the step-by-step shift operations.

Each shiftable SRAM cell includes a conventional SRAM cell, a CMOS transmission gate controlled by $\varphi_1$ as the inter-cell switch, and two NMOS switches controlled by $\varphi_2$ and $\varphi_{2d}$ as intra-cell switches.

The shift operation between adjacent SRAM cells consists of three phases, shown as a shift-right function in Fig. 3. In phase 1, the intra-cell switches controlled by $\varphi_2$ and $\varphi_{2d}$ are turned off and the inter-cell switch controlled by $\varphi_1$ is turned on. The remnant charge at node X will drive the two inverters to generate a path from the left cell to the right. In phase 2 and phase 3, the intra-cell switch controlled by $\varphi_2$ and $\varphi_{2d}$ are turned on one by one with other switches remaining off, so that each SRAM cell forms a closed loop to stabilize its datum.

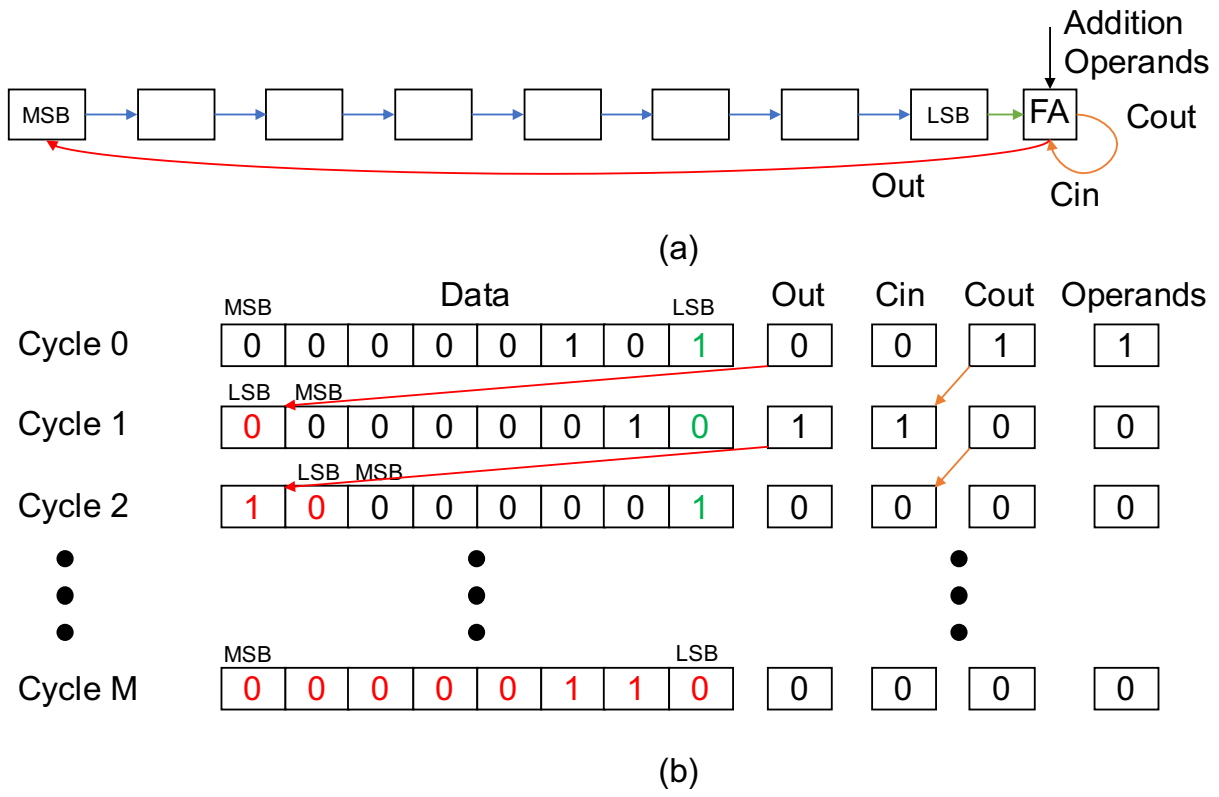

Fig. 4. Multi-bit addition with 1-bit full adder (FA).

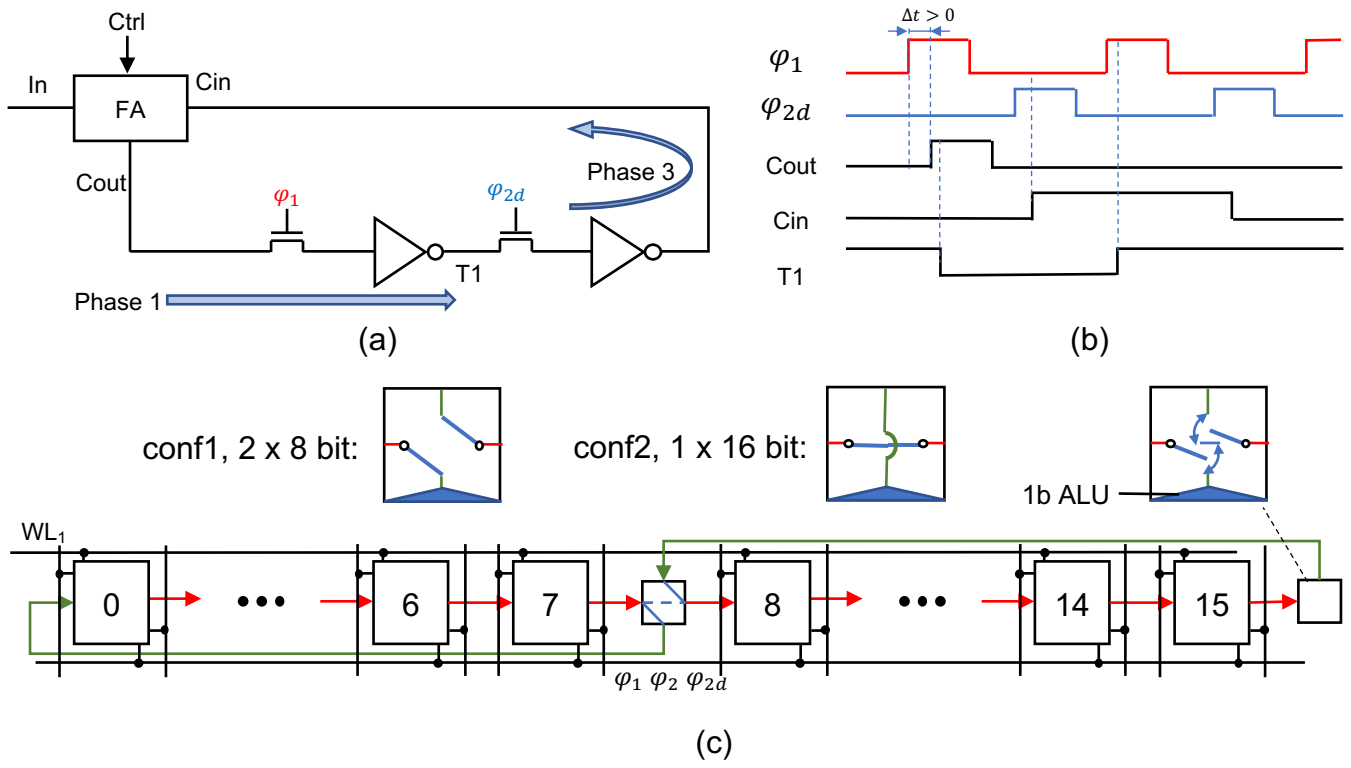

Fig. 5. Full adder with carry propagation: (a) circuit; (b) timing diagram, and (c) multi-word configuration route unit.

The timing of the control signals is shown in Fig. 3 (b). It could be conveniently generated by two-phase non-overlapping clock and a delayer. The control signals $\varphi_1$ and $\varphi_2$ are non-overlapping to avoid the data loss caused by the simultaneous turning-on of the switches. The control signal $\varphi_{2d}$ is set to $\varphi_2$ with a slight delay to provide sufficient time for data restoration in phase 2. The delay circuit could be simply realized by two serially connected inverters.

Fig. 7 shows the transient waveforms during the shift operations, in which the three phase control signals and the internal nodes of the four SRAM cells in a row are included.

## C. One-Bit ALU for In-Row Computing Capabilities

Based on the above in-memory shifter, by adding a 1-bit ALU to the end of each row (between LSB and MSB), the high parallel memory computing operation could be realized, as showcased in Fig. 4(a) with a full-adder (FA) example. For a $q$-bit datum stored in the row, after $q$ right-shift cycles along with the 1-bit FA, the external add operand will be added to the row and the data in this row are restored. An example of $q$=8 is shown in Fig. 4(b). Since the long interconnecting wires between the memory cells and ALU may result in large parasitic resistance and capacitance, folding each row back to form an evenly distributed loop is effective to limit the maximum distance to ~2x of the FAST SRAM design, shown as Fig. 6(b).

It is also noted that, during the multi-bit add, the FA carry bit needs to be stored temporarily. The circuit diagram of passing the carry bit to the next stage is shown in Fig. 5(a). In phase 1, FA calculates 1-bit addition and outputs the sum and carry-out bit. The switch $\varphi_1$ is turned on while $\varphi_{2d}$ is turned off to store the carry bit on the node T1. In phase 3, the carry will be transmitted through the switch $\varphi_{2d}$, which will be used as the input carry of the next stage. An example is shown in Fig. 5. (b) to showcase the workflow. In addition, we propose a bit-width reconfigure method, shown as Fig. 5(c) using 16 cells as an example. When we need to connect two low-width words as a wide-width word, the routing unit will connect the shift line of these two words. In this case, two ALU will be cascaded.

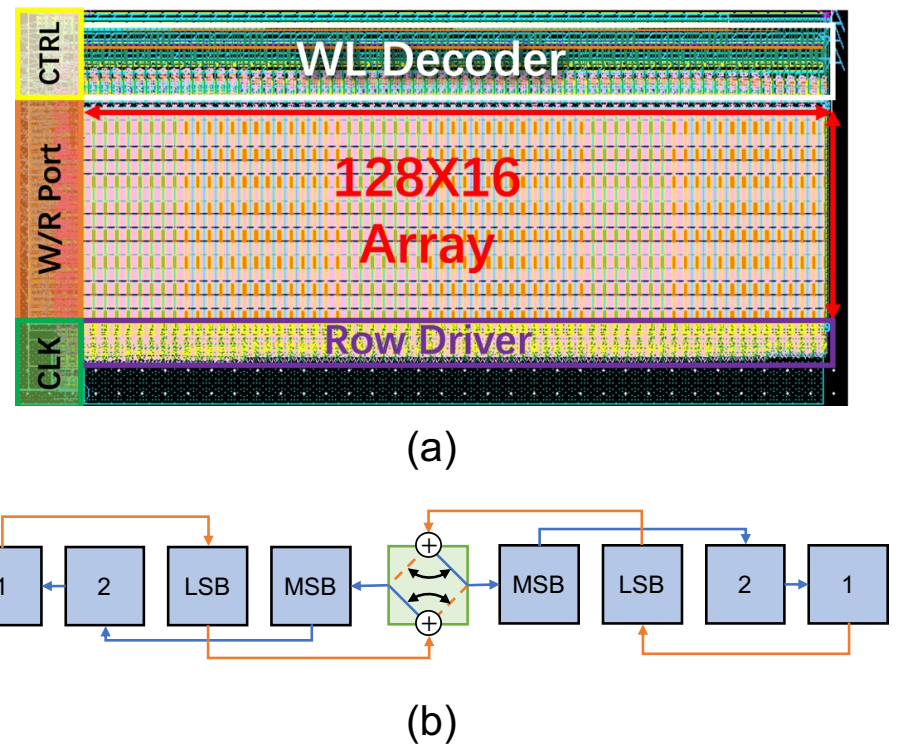

Fig. 6. FAST SRAM (a) layout, and (b) in-row cross structure.

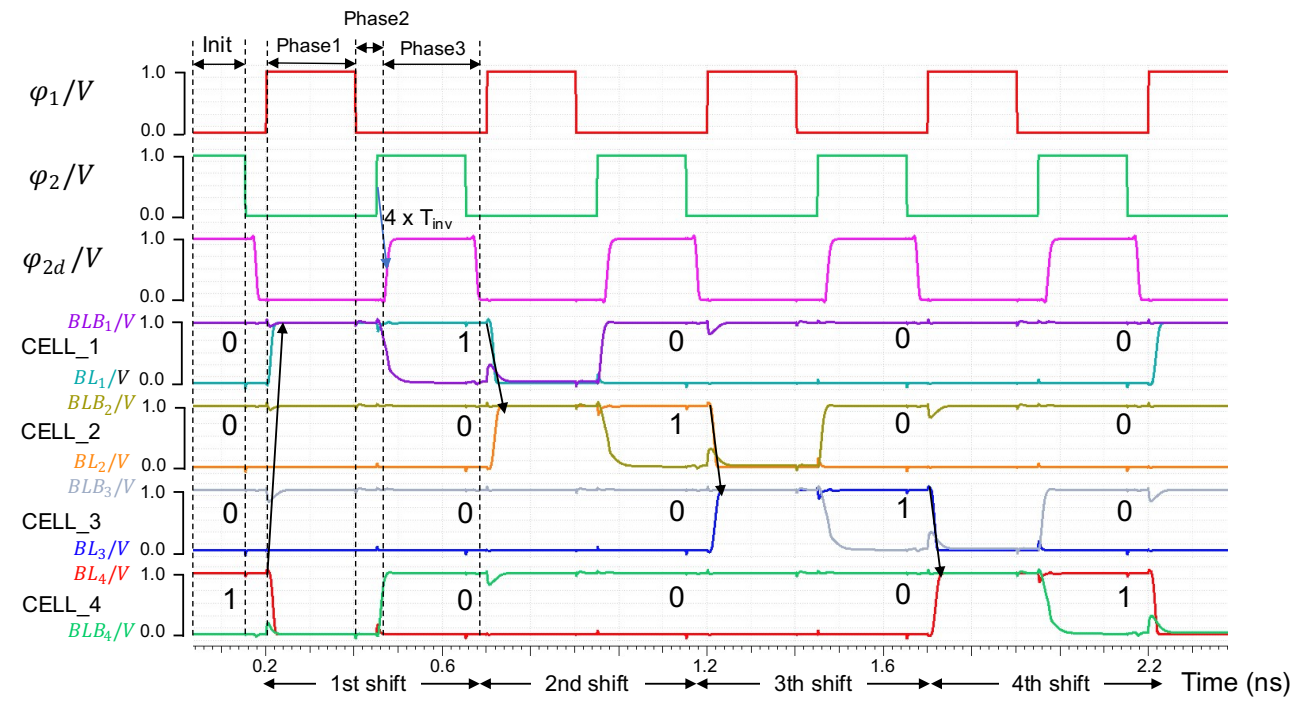

Fig. 7. Transient waveforms of shift operation.

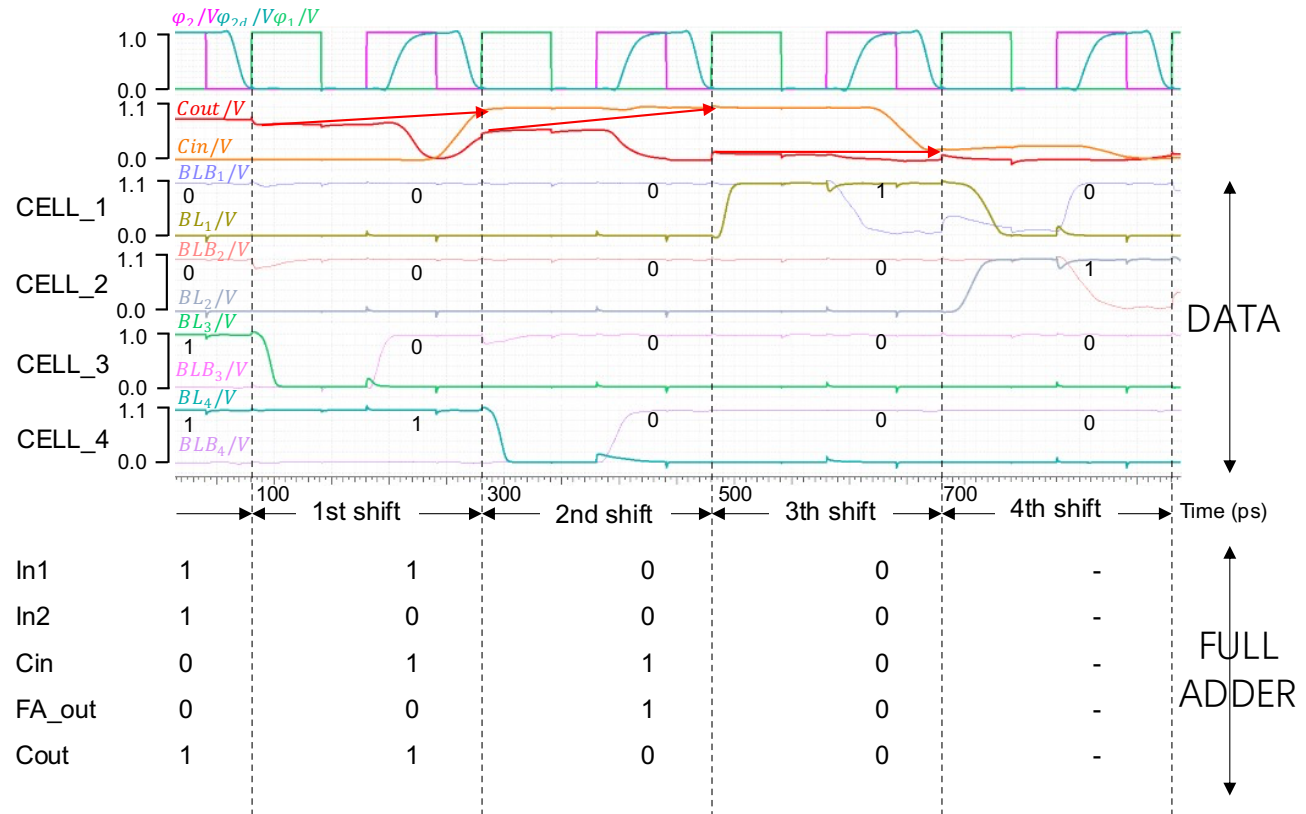

Fig. 8. Transient waveforms of 4-bit add with a 1-bit full adder.

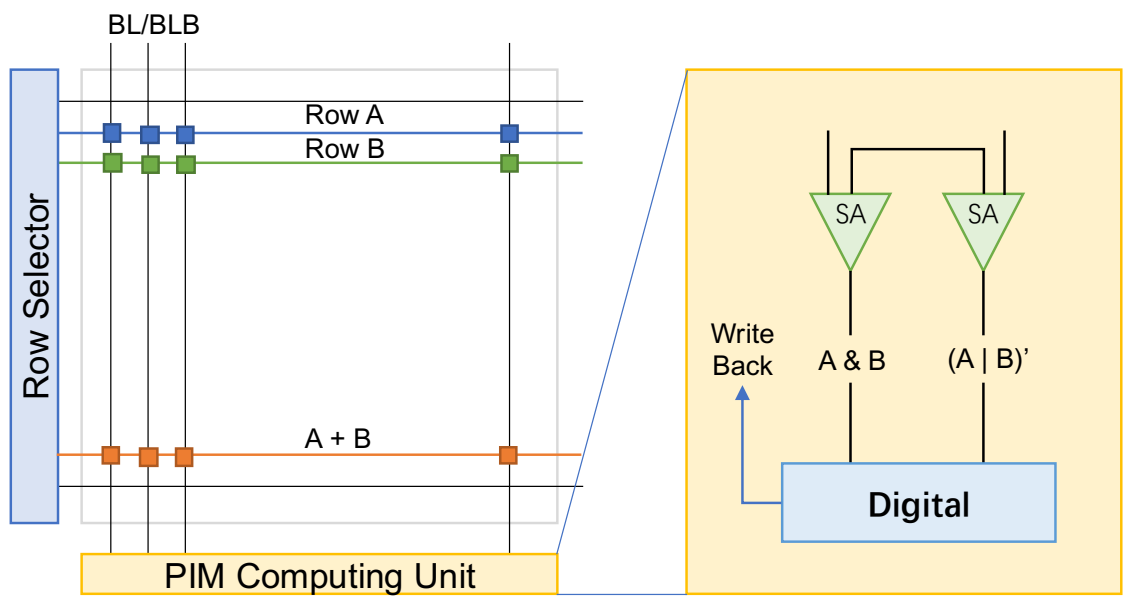

Fig. 9. The baseline of a fully-digital near-memory computing architecture.

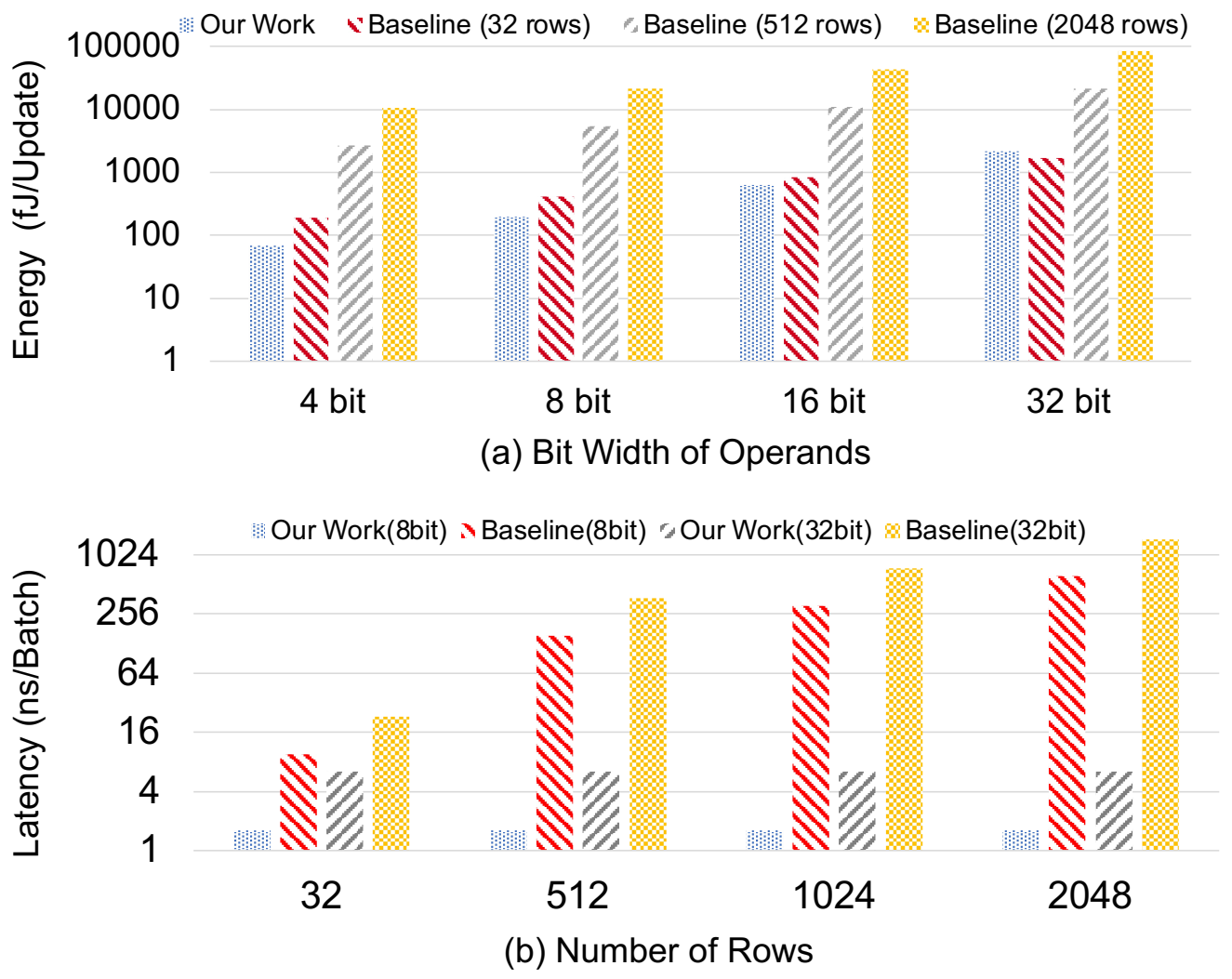

Fig. 10. Energy and latency comparison with different bit width.

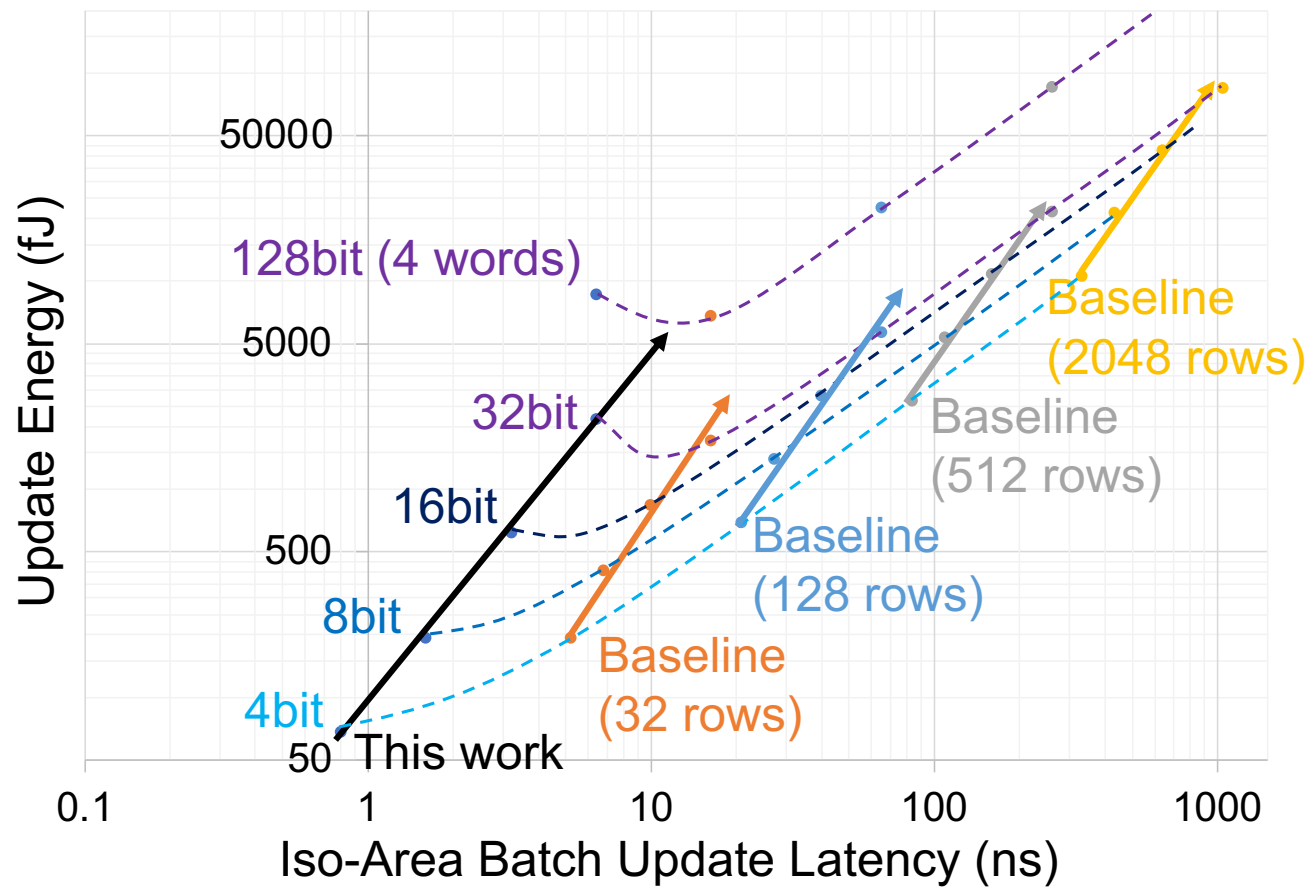

Fig. 11. Latency of batch update and energy efficiency at different bit width, which is normalized into the same area.

Fig. 8 shows the transient waveforms of the shift-based add operations, in which the phase control signals and the internal nodes of the four SRAM cells in a row are provided.

## III. Benchmarking And Discussion

This section evaluates the proposed FAST SRAM in terms of array-level power and latency, along with application testbenches. Simulation results of circuit transient waveforms, chip layout, etc. are included.

### *A. Chip Layout Design and Simulation Setting*

To evaluate the performance of the proposed FAST SRAM, a chip has been designed in a 65nm CMOS technology, as shown in the die photograph in Fig. 6. Post-layout parasitic exaction has been carried out for more accurate SPICE simulations. The supply voltage is set to 1.0 V.

For comparison purposes, a fully-digital near-memory computing architecture is chosen as the baseline design, shown as Fig. 9. The baseline is the general-purpose SRAM assisted with custom digital logic circuits designed by the digital flow based on a standard cell library. This baseline is built with the

TABLE I
Comparison Between SRAM Cache and Processing in Memory

| | FAST SRAM | SRAM | Digital |
|---|---|---|---|
| Cell Structure | 10T | 6T | 20T |
| Write Energy | 76.2 fJ/bit | 72.4 fJ/bit | 219.7 fJ/bit |
| Read Energy | 74.8 fJ/bit | 68.4 fJ/bit | / |
| Access Time | 0.94 ns | 0.94 ns | 0.09 ns |
| Calc. Energy * | 0.38 pJ/OP | / | 2.09 pJ/OP |
| Calc. Time * | 0.025 ns/OP | / | 0.68 ns/OP |

* OP: 16-bit addition with data write-back to the FAST SRAM in 128-row parallelism

same function as the FAST SRAM. For general benchmarks, the proposed architecture and the baseline architecture are both based on the conventional 6T SRAM structure. While simulating the performance and costs during the parallel update, we collect the energy consumption of each word update, and the latency of updating the whole array, i.e., the batch update latency.

### *B. Simulation Result*

*Energy Efficiency.* Fig. 10 (a) shows the energy consumption comparison. When the number of rows is greater than 2 times of the bit width, the proposed FAST SRAM scheme has higher energy efficiency. As mentioned above, this advantage comes from a shorter critical charging and discharging path of a memory access. When the bit width is much less than the number of rows (which is commonly adopted in general SRAM design to reduce the costs of the the peripherals), the energy saving is significant. For example, the energy efficiency could be 4.4x higher than the baseline with 8-bit bit width and 512 rows (here the number of rows is 64x of the bit width).

*Latency.* Fig. 10 (b) shows the latency comparison, in which the proposed FAST SRAM scheme shows hundreds of times speedup. The reason behind this advantage is straightforward: the latency of baseline depends on the number of rows in the array to carry out the operations row by row, while the proposed FAST SRAM support full-concurrency operations on all the rows and the latency depends on the bit width. When the number of rows is larger compared to the bit width, the latency advantage of high parallelism becomes more significant.

Fig. 11 provides more simulation results to highlight the trend of the latency and energy consumption of the proposed scheme under different bit width and number of rows.

### *C. Performance Benchmarking*

FAST technique provides a new circuit functionality to support a category of parallel memory access with light logic operations and in-place updates. On the one hand, FAST SRAM could be useful for conventional applications such as database indexing, in-memory search or sorting, etc. Essentially, as conventional SRAM does not support access to multiple rows at a time, FAST SRAM provides a new potential to deal with other parallel data-update operations. On the other hand, since FAST can handle parallel search and update operations, it is also useful in emerging graph computing applications.

As an example, table I compares the latency and energy performance of addition and update in FAST SRAM under a configuration of 128 rows in the benchmark. Compared with the fully-digital computing architecture, our design shows 5.5x energy saving and 27.2x speedup. The proposed FAST SRAM architecture benefits from the concurrent operations.

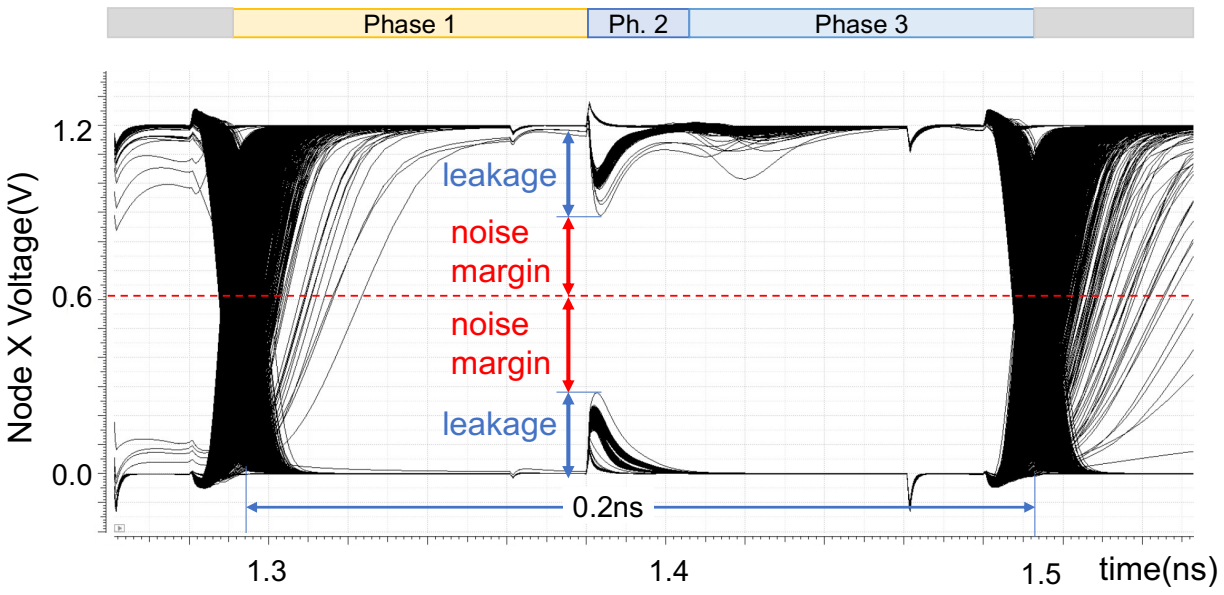

Fig. 12. Noise tolerance and stability analysis.

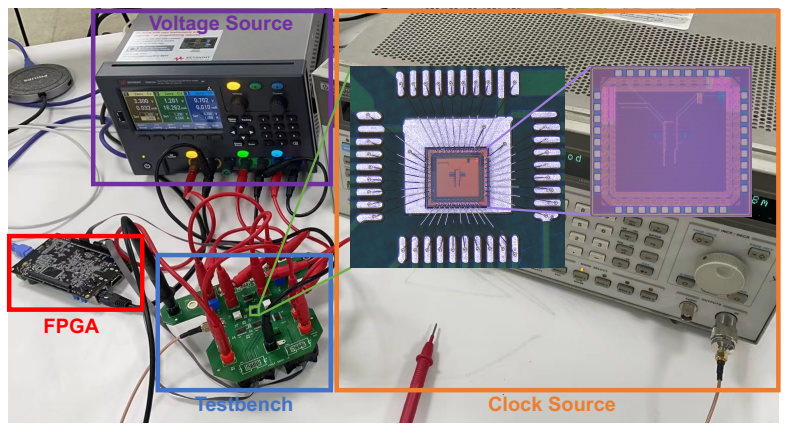

| | 500MHz | 800MHz | 1GHz | 1.2GHz |
|---|---|---|---|---|
| 1.2 V | Failed | Success | Success | Success |
| 1.1 V | Failed | Success | Success | Failed |
| 1.0 V | Failed | Success | Failed | Failed |
| 0.9 V | Failed | Failed | Failed | Failed |

Fig. 13. Test environment and shmoo plot for the region of FAST SRAM.

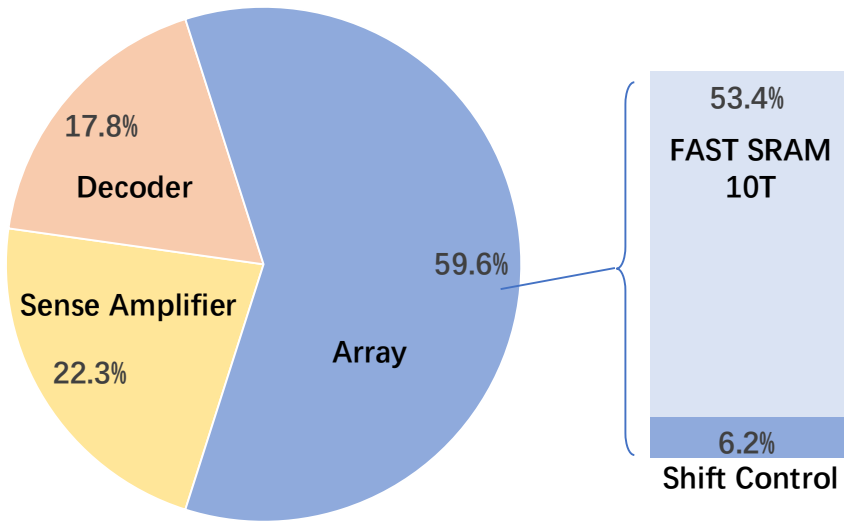

Fig. 14. Area breakdown on the die.

### *D. Test Results and Validation*

Since data transfer between SRAM cells is a dynamic logic, the noise margin is critical. In phase 2, the switches $\varphi_{2d}$ and $\varphi_1$ will be off. Therefore, the charge stored in the start point of the disconnected inverters loop in FAST SRAM will leak slowly, as shown in Fig. 12. Monte Carlo simulation validates the eye pattern of in-row shift with different cells. There is still a 300mV noise margin in the worst case.

A tested macro on SMIC 55nm is verified. The test environment and shmoo plot are shown in Fig. 13.

### *E. Overheads, More Discussions, and Future Works.*

In the experimental chip design, the proposed FAST SRAM architecture adopts ten transistors per cell, including six original SRAM cell transistors and four switch transistors. This extra transistor count brings about 70% area overhead on cell level in our design. The area overhead of shift control signal generation is only about 10% in a 16-column scenario. Fig. 14 illustrate the area breakdown of a 128-row FAST SRAM die. Considering the peripherals, the FAST SRAM takes about 41.7% more area compared with the general-purpose SRAM.

It is also noted that, the proposed SRAM subarray could also be used as a general cache, especially for data-intensive applications such as multimedia processing and encryption. In addition, it can also realize more complex functions by replacing the 1-bit full adder into other 1-bit operation units. This architecture could serve as a data in-situ update accelerator with high energy efficiency for inference acceleration, database index search and other applications with high-concurrency row-by-row operations. Future work may also consider a reconfigurable design to deal with more versatile calculations such as float point adder or integer multiplier.

## IV. Conclusion

This paper has proposed a novel memory and architecture, namely FAST, which is capable of dealing with high-concurrency row-wise memory operations. A chip designed in 65nm CMOS technology has been showcased to demonstrate its efficiency in parallel memory access and data update operations. The overhead of the proposed design is mainly the area overhead. Future work that further harnesses the parallelism, flexibility, reconfigurability could be meaningful in data-intensive applications where high-concurrent memory access is the performance bottleneck.